\documentstyle[tighten,aps]{revtex}

\begin{document}

\draft
\twocolumn

\title{ On Non-- Efficiency of Quantum Computer}

\author{Robert Alicki}

\address{Institute of Theoretical Physics and Astrophysics, University
of Gda\'nsk, Wita Stwosza 57, PL 80-952 Gda\'nsk, Poland}

\date{\today}
\maketitle

\begin{abstract}

Let $E_c$ be the energy used by the quantum computer
to perform the computation, $t_c$ the total computation time including
the preparation of an input and the measurement of an output state and
${\cal C}$ the complexity of the problem defined as a minimal number
of logical steps needed to solve it. We advocate as a plausible
hypothesis
a previously proposed inequality motivated by the
Heisenberg energy-time uncertainty principle which has the form $E_c t_c
>>
\hbar {\cal C}$. This hypothesis is supported by the
following explicit examples of quantum operations and computations:
preparation of an input $n$ $q$-bit state, two Hamiltonian versions
of the Grover's search algorithm, a model of a "quantum telephone
directory",
a quantum-optical device which can factorize any number and a network
used
in Shor's algorithm.
\end{abstract}

\section{Introduction}

The standard description of a quantum computer is the following. We have
a
quantum system with $N=2^n$ orthogonal states
(computational basis) which can store  $n= \log_2 N $ bits of
information
typically realized as a collection of $n$ 2-level subsystems ($q$-bits).

The process of computation is divided into three stages, first
an initial (input) state is prepared,
then a quantum algorithm is performed which is realized as a sequence
of ${\cal G}$ unitary transformation called quantum gates and finally,
the output
state is measured. The number ${\cal G}$ of involved quantum gates is
called
sometimes a quantum complexity and one assumes, in analogy to classical
digital computers, that the physical time needed to achieve a given
task is proportional to ${\cal G}$. If ${\cal G}$ is polynomial in $n$
the
algorithm is called efficient .
The remarkable result of Shor [1] had an enormous impact
on the whole field of quantum information and quantum computing [2]. He
constructed a quantum algorithm as a sequence of polynomial in $\log_2
N$
number of unitary transformations which factorizes numbers smaller than
$N$
into primes. It is known that the logical complexity $\cal C$ of the
problem
measured by a number of logical steps needed to solve it
grows exponentially with $n =\log_2 N$ for the case of factorization.
The possibility of efficient factorization of large numbers by quantum
computers must frighten the experts responsible for the safety of
information
transmission.
\par
The reasoning outlined above possesses several drawbacks. First of all
the idea that the physical time of computation $t_c$ is proportional to
the complexity
$$
t_c \sim {\cal C}\ {\rm or}\ t_c \sim {\cal G}
\eqno(1)
$$
is only true for the existing digital computers which are ensembles of
controlled bistable elements which correspond to Boolean logical values
0,1
and can literary mimic logical operations.
Other theoretically conceivable classical computers like, for example,
ballistic computer of Fredkin and Toffoli [3](hard spheres colliding
with each other and with fixed reflective barriers) need not satisfy
(1).
By rescaling masses,  distances and initial velocities we can have in
principle for a fixed $N$ arbitrarily short $t_c$ and arbitrarily
low energy store needed. Obviously, atomic structure puts limits on the
rescaling and classical chaos together with friction make the idea of
ballistic computers impractical.
\par
The number of quantum gates ${\cal G}$ which form a quantum
algorithm is not unique and does not determine the real physical
time of computation. The very idea of quantum gate is not natural in
quantum
mechanical context. Quantum evolution is continuous in time and governed
by
a (possibly time-dependent) Hamiltonian which describes the energy of a
system. In particular by increasing the energy level spacing of
the computer's Hamiltonian we can speed up its time evolution. Hence the

physical efficiency of computation should be given in terms of the
product of
a characteristic energy and time ("action parameter") rather than the
computation time alone.  A dynamical character and finite duration of
the
input state preparation procedure and the output state measurement
process
are also important. As we shall see studying concrete examples there is
no
clear separation between the preparation-measurement processes and the
action of quantum algorithm.
\par
As many authors noticed the decoherence effects due to the interaction
with
an environment are the main practical obstacles for the operation of
quantum
computer. As the decoherence typically grows in a nonlinear way with the

energy level spacing this can produce an optimal splitting of the action

parameter into time and energy. This problem is studied in details in
[4] and will be not discussed here.

\section{Heisenberg energy-time relations}

One of the interpretations of the Heisenberg energy-time uncertainty
relation
$$
\Delta E\cdot\Delta t \geq \hbar
\eqno(2)
$$
is that a quantum state with spread in energy $\Delta E$ takes time at
least
$\Delta t = \pi \hbar /2\Delta E$ to evolve to an orthogonal and hence
distinguishable state [5]. Hence, it is quite natural to investigate the

quantum mechanical limitations on the action parameter $E_c\cdot t_c$
where $E_c$ is an energy store needed for the quantum computation and
$t_c$ is
a total computation time including the preparation of an input state and

the measurement of an output. Following [6] we can propose the
inequality
of the form
$$
E_c\cdot t_c >> \hbar{\cal C}_q
\eqno(3)
$$
where ${\cal C}_q$ is a not yet defined "quantum complexity" of the
problem.
The "much larger" symbol in (3) takes into account the probabilistic
character
of quantum algorithms which should run a certain number of times to
achieve
a given level of confidence. Although is it not formulated explicitly in

the form
of the inequality (3) the general believe among the experts in quantum
computation
is that the quantum complexity is equal to the minimal number of quantum

gates
$$
{\cal C}_q = {\cal G}\ .
\eqno(4)
$$
In [6] the following hypothesis supported by a single
example has been proposed
$$
{\cal C}_q \equiv {\cal C}
\eqno(5)
$$
what means essentially that there is no {\it quantum } complexity.
In particular,
if the problem needs an exponential in the input bit size number of
logical steps the quantum computer needs also an exponential time or
an exponential energy store. It is still better than for the existing
digital
computers which need both but nevertheless the quantum computation is
not
practically efficient.

\section{Examples}

In order to advocate the hypothesis (5) we discuss several examples
of quantum algorithms or their essential parts.

\subsection{ State preparation}

Consider $n$ two level system ($q$-bits) with a standard computational
basis which consists of products of two $q$-bit states $|0>$ and $|1>$.
To prepare an arbitrary
input state from this basis starting from a certain fixed initial state
we need to rotate some of the 2-level system
in a fixed time $t_c$ to the orthogonal states. Therefore according to
the
presented interpretation of (1) we need on the average the energy
$E_c = (n/2) \pi\hbar/2t_c$ what gives in this case
$$
{\cal C}_q \approx n\ .
\eqno(6)
$$
The similar result can be obtained for the quantum measurement of an
output
state.
As writing or reading of $n$-bit messages involves $n$ logical steps we
have
here ${\cal C}_q = {\cal C}$

\subsection{Grover's search algorithms}

We shall analyze two different Hamiltonian realizations of the Grover's
search algorithm [7]. In the original formulation every element of the
database is represented by a state of the standard computational basis
of the $n$ $q$-bit system. One of this states -- a searched one-- is
denoted
by $|x>$ and we have also a certain standard initial state $|in>$ which
is usually a uniform superposition of computational basis $\{|\phi_j>;
j= 1,2,...,N=2^n\}$
$$
|in> = (N)^{-1/2}\sum_{j=1}^N |\phi_j>,\ \ <s|in> = N^{-1/2}\ .
\eqno(7)
$$
These two states are hidden in the dynamics of the system described
either
by the repeated unitary transformations or the continuous time
Hamiltonian
evolution.
The first Hamiltonian proposed in [8] has form
$$
{\bf H}_1 = E (|x><x| + |in><in|)
\eqno(8)
$$
where $E$ is an energy scale. The Hamiltonian acts essentially on the
two
dimensional subspace of the Hilbert space and the corresponding energy
difference
is equal to
$$
\hbar\omega_N = E N^{-1/2}\ .
\eqno(9)
$$
One can easily compute that we need a time
$$
t_N = \pi N^{1/2}/2E
\eqno(10)
$$
to reach from the initial state $|in>$ the searched one $|x>$. So the
quantum
complexity of this stage of searching is of the order $\hbar\omega_N
t_N\approx 1$. Adding preparation
and measurement processes we obtain
$$
{\cal C}_q\approx n\ .
\eqno(11)
$$
The second Hamiltonian used with the different rescaling in [9] reads
$$
{\bf H}_2 = i E (|x><in| - |in><x|)\ .
\eqno(12)
$$
Again the problem is essentially two dimensional with the energy
difference
$$
\hbar\omega_N = 2E + o(1/N)\ .
\eqno(12)
$$
The time needed to reach the state $|x>$ can be estimated by
$$
t_N = \pi/4E + o(1/N)\ .
\eqno(13)
$$
Here again adding state preparation and measurement we obtain (11).
\par
In the literature on the Grover's algorithm it is claimed that its
quantum
complexity is $\sqrt{N}$ which is compared with the classical complexity

of the problem claimed to be equal to $N/2$. Both statements are
incorrect.
We have just
computed the quantum complexity of the problem equal to $n=\log_2 N$.
The classical analogon of Grover's search algorithm is not finding an
item
in a randomly ordered phone book but rather a search for a one heavier
ball among $N$ otherwise indistinguishable ones. The later problem
can be solved in $\log_2 N$ steps.

\subsection {Quantum telephone directory}

We discuss now a true quantum analog of a random telephone directory.
It is again a $n$ $q$-bit system with a computational basis $\{\phi_j;
j= 1,2,...,N= 2^n\}$. We fix an initial state to be $\phi_1$ and
propose the following time dependent Hamiltonian
$$
{\bf H}(t) = {\bf H}_0 + {\bf V}\cos (\Omega t)
\eqno(14)
$$
where
$$
{\bf H}_0 = \sum_{j=1}^N E_j |\phi_j><\phi_j|
\eqno(15)
$$
${\bf V}$ is a (randomly chosen) weak perturbation and $\Omega$ is a
tunable
frequency. The energies $E_j\geq 0$ are not degenerated and provide
labels
for the states $\phi_j$ (we put $E_1 = 0$). To find a state labeled by
$E_j$ we tune the frequency to the value $\Omega_j = E_j/\hbar$.
The time dependent first-order perturbation calculus [10] gives us the
probability of excitation of the state $\phi_k$
$$
p_k(t)= 2|<\phi_1|{\bf V}|\phi_k>|^2\
{\sin^2 \bigl\{{1\over 2}
\bigl(E_k - E_j) t/\hbar\bigr\}\over\bigl(E_k - E_j)^2}\ .
\eqno(16)
$$
It follows from the formula (16) that we have to wait
for a time at least of the order
$$
t_j \approx {\hbar\over |E_j - E_k|}
\eqno(17)
$$
to be sure that the searched state $\phi_j$ has
been prepared with a much larger probability than the other neighboring
state $\phi_k$. Therefore, on the average we obtain the computation time

$$
t_c \approx \hbar N/E_{max}
\eqno(18)
$$
where $E_{max}={\rm max} \{E_j\}$.  Then as $E_c\approx E_{max}$ we see
again that the quantum complexity coincides with the classical one.

\subsection{ Quantum device factorizing numbers}

This model has been introduced in [6] but we briefly discuss it again
for
the sake of completeness. In fact this model is very similar to the
previous one.
\par
A resonant cavity supports radiation modes with the frequencies being
the logarithms of prime numbers times a fixed frequency unit $\omega$
$$
\omega_q = \omega\log q\ ,\  q = 2,3,5,7,11,13,...
\eqno(19)
$$
The second quantization Hamiltonian of the electromagnetic field
$$
{\bf H} = \hbar \omega\sum_q (\log q)\  a^+_q a_q
\eqno(20)
$$
has nondegenerated eigenvalues being proportional to the logarithms
of all natural numbers
$$
{\bf H}\psi_N = E_N \psi_N\ ,\ E_N = \hbar\omega\log N,\  N = 1,2,3,...
\eqno(21)
$$
The structure of $\psi_N$ reveals the factorization
of $N$ into prime numbers
$$
\psi_N \sim (a^+_{q_1})^{m_1}(a^+_{q_2})^{m_2}\cdots
(a^+_{q_r})^{m_r}\psi_1
\eqno(22)
$$
where
$$
N = (q_1)^{m_1}(q_2)^{m_2}\cdots(q_r)^{m_r}
\eqno(23)
$$
and $\psi_1$ is a vacuum state. The eq.(22) means that we have
$m_1$ photons of the frequency $\omega\log q_1$, $m_2$ photons of the
frequency
$\omega\log q_2$, ..., and $m_r$ photons of the frequency $\omega\log
q_r$.
Therefore, transferring
a given energy portion $\hbar\omega\log N$ to the empty cavity and then
opening the cavity and counting photons in different modes
we obtain the factorization of $N$.
It can be done similarly to the previous example perturbing the system
periodically in time with a tunable frequency $\Omega$
and selecting $\Omega = \omega\log N$.
\par
As the energy level spacing around $E_N = \hbar\omega
\log N$  is $\delta E_N \approx
\hbar\omega/N$ it follows from the analogue of the formula (16) that we
have
to wait  for a time at least of the order
$$
t_c \approx N \omega^{-1}
\eqno(24)
$$
to select a proper state $\psi_N$. The energy used equals
$E_c= \hbar\omega\log N$
and once again the quantum complexity essentially coincides with
the classical one.

\subsection{Shor's algorithm}

The Shor's factorization algorithm is rather complicated but for our
purposes
we need only a
part of it -- the so-called phase shift computation. We follow here the
simple presentation in [11]. The phase shift computation consists in
applying to the initial state of $n$ $q$-bit system
$$
|in> = 2^{-n/2}(|0>+|1>)\otimes\cdots\otimes(|0>+|1>)
\eqno(25)
$$
a $n$-gates unitary operation
$$
{\bf U} = U^{2^0}\otimes U^{2^1} \otimes\cdots\otimes U^{2^n}
\eqno(26)
$$
where
$$
U|0> = |0>\ ,\ \ U|1> = e^{-i\alpha}|1>\ ,\ \alpha\in[0, 2\pi)
\eqno(27)
$$
The unitary ${\bf U}$ can be realized as $\exp\{-i{\bf H}t_n/\hbar\}$
with the following  Hamiltonian
$$
{\bf H} = \hbar\omega\sum_{k=0}^{n-1} 2^k (|1><1|)_k
\eqno(28)
$$
acting for a time $t_n = \alpha/\omega$. The averaged energy in the
state
$|in>$ grows exponentially with $n$ and is given by
$$
{\bar E}_n = <in|{\bf H}|in> =
 (\hbar\omega/2)\sum_{k=0}^{n-1} 2^k = \hbar\omega (2^{n-1} -1/2)\ .
\eqno(29)
$$
For the total factorization procedure $t_c >> t_n$ and $E_c >> {\bar
E}_n$.
Taking for $\alpha$ its average value $\pi$ we have
$ E_c t_c >>\hbar\,2^n = \hbar N$ in agreement with the hypothesis (5).

\section {Conclusion}
Although the general proof would be very desirable
the analysis of the presented examples, in particular the powerful
Shor's
algorithm,
provides a strong evidence for the Heisenberg-like bound (3)(5)
on the efficiency of quantum computations.

\acknowledgments
The author thanks Micha\l, Pawe\l\ and Ryszard Horodecki for
discussions.
The work is supported by the Grant KBN PB/273/PO3/99/16.


\begin{references}

\bibitem{[1] } P. Shor, in {\it Proceedings of the 35th Annual Symposium

on Foundations of Computer Science}, edited by S. Goldwasser (IEEE
Press, New
York, 1994), pp. 56--65.

\bibitem{[2] } H.-K. Lo, S. Popescu, and T. Spiller (eds.)
{\it Introduction to Quantum Computation and Information}, (World
Scientific,
Singapore,  1998).

\bibitem{[3] } C. H. Bennet , Int.J.Theor.Phys. {\bf 21}, 905 (1982)

\bibitem{[4] } R. Alicki, M. Horodecki, P. Horodecki, and R. Horodecki
(in preparation)

\bibitem{[5] } Y. Aharonov and D. Bohm, Phys.Rev. {\bf 122}, 1649 (1961)

\bibitem{[6] } R. Alicki, e-print quant-ph/0006018

\bibitem{[7] } L.K. Grover, Phys.Rev.Lett. {\bf 79}, 325 (1997)

\bibitem{[8] } E. Farhi and S. Gutmann, Phys.Rev. A {\bf 57}, 2403
(1998)

\bibitem{[9] } M. Mussinger, A. Delgado and G. Alber, e-print
quant-ph/0003141

\bibitem{[10] } A. Messiah, {\it Quantum Mechanics} , (North-Holland,
Amsterdam, 1962).

\bibitem{[11] } R. Cleve et.al., e-print quant/9903061

\end{references}
\end{document}